\documentclass[pre,twocolumn,showpacs,floatfix]{revtex4}
 \usepackage{graphicx}
 \usepackage{epstopdf}
 \usepackage{dcolumn}
 \usepackage{color}
 \usepackage{bm}
 \usepackage{verbatim}
 \usepackage{multirow}
 \usepackage{amsmath,amssymb,graphicx}
 \usepackage{epsfig}
 \usepackage{enumerate}
 \usepackage{array}
 \usepackage{multirow}
\begin{document}
\title{Brownian motion of classical spins : Anomalous dissipation and generalized Langevin equation}
\author{Malay Bandyopadhyay$^1$ and A. M. Jayannavar$^2$}
\affiliation{1. School of Basic Sciences, Indian Institue of Technology Bhubaneswar, Bhubaneswar, India 751007\\
2. Institute of Physics, Sachivalaya Marg, Sainik School PO, Bhubaneswar, India, 751005.}
\begin{abstract}
 In this work, we analyze the relaxation of a classical spin interacting with a heat bath, starting from the fully dynamical Hamiltonian description. An analogous problem in the framework of generalized Langevin equation (GLE) with anomalous dissipation is analyzed in details. The Fokker-Planck equation corresponding to GLE is derived and the concept of equilibrium probability distribution is analyzed. In this process we have identified few difficulties to obtain equilibrium distribution for the non-Markovian case and the remedy to overcome this difficulty is also discussed.
\end{abstract}
\maketitle
\section{Introduction}
The most well known and well studied method of Brownian motion is the Langevin dynamics which has a long history. \cite{chandrasekhar,kubo,coffey}.
One can model the dynamics of relaxation of an ensemble of
interacting particles evolving toward thermal equilibrium by incorporating fluctuation and concomittant
dissipation forces in the Hamiltonian equations of motion. The existence of a relation
 between the stochastic and dissipative Langevin terms
arises from the fluctuation-dissipation theorem (FDT)\cite{chandrasekhar,kubo,coffey} to
maintain the requirement that in thermal equilibrium the
population of states in the phase space is given by the Gibbs
distribution. A similar Langevin approach can be adopted to the investigation
of the relaxation of a single spin interacting with a heat bath or an interacting spin ensemble by introducing proper stochastic and dissipation terms in the spin equation of motion. This was proposed by Brown in ref. \cite{brown}. A single spin interacting with a heat bath or an ensemble of interacting spins are driven toward equilibrium by suitably chosen fluctuation and dissipation terms. The basic requirements of choosing fluctuation and dissipation terms is to satisfy the fluctuation-dissipation theorem. The treatment also
makes it possible to model equilibrium thermal fluctuations
associated with the dynamics of either individual atomic spins
or the magnetic moments of small particles.\cite{garcia,chantrell}. Thus, the relaxation phenomena of spin plays a crucial and important role in condensed matter. Mainly this will help us to understand the underlying physical processes associated with the lineshapes of NMR, ESR etc. Very recently studies have been carried out in the context of fluctuation theorem and entropy production of a large macrospin with fluctuating magnitude and direction \cite{debc1,debc2}. The complicated motion of the spin due to its interaction with the surrounding medium can be modelled with the introduction of a stochastic magnetic field $\vec{H}(t)$. Assuming Gaussian noise properties for $\vec{H}(t)$, one usually arrives at the well known Bloch equation\cite{abragam,slichter,kubo1,kubo2}. But, this model  unable to produce correct equilibrium value $\chi \vec{H}_0$ in the presence of an external magnetic field $\vec{H}_0$ with magnetic susceptibility $\chi$ for spin. Rather, the average magnetic moment relaxes to zero which is nothing but the infinitely high temperature value. Kubo and Hashitsume \cite{kubo3} introduced the concomitant friction term $-\eta \frac{d\vec{M}}{dt}$ to overcome the above mentioned problem which resembles the Landau-Lifshitz equation with nonlinear friction force \cite{landau}.\\
Several attempts were made to derive the nonlinear Langevin equations for a classical spin interacting with a heat bath\cite{mahato1,mahato2} from the microscopic Hamiltonian description which is consistent with fluctuation-dissipation theorem \cite{kawabata,seshadri}. Seshadri and Lindenberg derived the nonlinear Langevin equation using full dynamical method by considering a test particle with a spin which interacts with a heat bath of other spins via Heisenberg type (i.e. Ruderman-Kittel interactions)\cite{seshadri}. This model enables one to derive the nonlinear Langevin equations of motion for different components of  spin of the test particle to the second order in spin-bath interaction and the relaxation parameters can be obtained directly from the model Hamiltonian. Unlike linear Bloch equation, their equations of motion of average spin components are also nonlinear and the Bloch equations can be recovered in the high temperature limit. Although, Seshadri and Lindenberg demanded that their equations automatically lead to the thermal equilibrium in the classical limit i.e. they obey the fluctuation-dissipation theorem. But, this is partially true. Equilibrium is only obtained in the direction of $\theta$, but the presence of a gyration term (first term in the r.h.s. of Eq. 2.21 (a) of \cite{seshadri} ) can generate a current along $\phi$ direction and it leads to a long time rotational motion with a certain frequency. Later, one of us considered a test particle with a spin and interacting with a heat bath consists of a set of generalized harmonic oscillators which are responsible for the relaxation of the test particle and the derived Langevin equations are again nonlinear, non-Markovian and they are consistent with the fluctuation-dissipation theorem \cite{jayannavar}.\\
In this paper we basically follow the method introduced in Ref. \cite{jayannavar}. Our basic focus is on the derivation of nonlinear Langevin equations for a spin interacting with a heat bath through momentum variables. Unlike the coordinate-coordinate coupling considered earlier in Ref. \cite{jayannavar}, our model preserves time reversal property of the system-reservoir Hamiltonian \cite{mahato1,mahato2}. In our case coordinate-momentum coupling scheme is extensively used in deriving the Langevin equation of our Brownian particle \cite{leggett,rfo1,rfo2}. Such kind of coupling is extensively used in the context of quantum tunneling in the presence of arbitrary linear dissipation by Leggett \cite{leggett} who termed this dissipation as anomalous dissipation and we follow his nomenclature. Further, this coordinate-momentum coupling is extensively used in the context of radiation reaction of a nonrelativistic electron interacting with quantum electrodynamic radiation field\cite{rfo1,rfo2} .  One important finding of our study is that the presence of anomalous dissipation does not alter much the basic features of the earlier study \cite{jayannavar}. We again obtain a nonlinear and non-Markovian Langevin equations which are consistent with the fluctuation-dissipation theorem.\\
The structure of our paper is as follows. In the next section we discuss our model Hamiltonian with momentum coupling. Then we derive the GLE corresponding to our Hamiltonian. In section III, we discuss about the corresponding Fokker-Planck equation and its stationary solution for both the Markovian and non-Markovian limits. Finally, we conclude in Sec. IV.\\
\section{Model Hamiltonian : Langevin Equations}
We consider a test particle with a spin interacting with a heat bath representing the environment consists of a set of harmonic oscillators. The spin of the test particle interacts with the heat bath through the momentum variables of the heat bath instead of usual position variables of the bath. An external magnetic field $H_0$ is taken along the z-direction. The closed system comprised of the test particle with its environment can be described by the Hamiltonian :
\begin{eqnarray}
\mathcal{H}= -H_0M_Z &+&\sum_{i=1}^{3}\sum_{\alpha=1}^{N}\Big\lbrack \frac{1}{2m_{\alpha,i}}(p_{\alpha,i}+ \lambda_{\alpha,i}M_i)^2 \nonumber \\
&+&\frac{1}{2}m_{\alpha,i}\omega_{\alpha,i}^2q_{\alpha,i}^2\Big\rbrack,
\end{eqnarray}
where $M_i; i=1,2,3$ describe the x, y, z components of the spin respectively, $p_{\alpha,i}, q_{\alpha,i}, \omega_{\alpha,i}$ and $m_{\alpha,i}$ represent canonical momentum, position, frequency and masses of the bath variables respectively. Here, $\lambda_{\alpha,i}$ represent the spin-bath coupling constants. Coupling induced renormalization of energy is compensated by the terms proportional to $M_i^2$. \\
As we know the classical equation of motion of any dynamical variable A obeyed the following evolution equation :
\begin{equation}
\dot{A}=\{A,\mathcal{H}\},
\end{equation}
where $\{-,-\}$ denotes Poisson bracket. The Poisson bracket relations among the spin variables are given by
\begin{equation}
\{M_i,M_j\}=\gamma\epsilon_{ijk}M_k,
\end{equation}
where $\gamma$ represents the product of the gyromagnetic ratio and the Bohr magneton of the spin and $\epsilon_{ijk}$ is the Levi-Civita symbol where $\epsilon_{ijk}=1$ for cyclic permutation of $i,j,$ and $k$. Moreover, the bath variables follow
\begin{equation}
\{q_i,p_j\}=\delta_{i,j}.
\end{equation}
Using Eqs. (2)-(4), we can obtain the equations of motion for the spin variables
\begin{eqnarray}
\frac{dM_z}{dt}&=&\gamma M_y\sum_{\alpha=1}^{N}\frac{\lambda_{\alpha,1}}{m_{\alpha,1}}p_{\alpha,1}-\gamma M_x\sum_{\alpha=1}^{N}\frac{\lambda_{\alpha,2}}{m_{\alpha,2}}p_{\alpha,2} \nonumber \\
&+&\gamma M_xM_y\sum_{\alpha=1}^{N}\frac{\lambda_{\alpha,1}^2}{m_{\alpha,1}}-\gamma M_yM_x\sum_{\alpha=1}^{N}\frac{\lambda_{\alpha,2}^2}{m_{\alpha,2}},
\end{eqnarray}

\begin{eqnarray}
\frac{dM_y}{dt}&=&-\gamma H_0M_x+\gamma M_x\sum_{\alpha=1}^{N}\frac{\lambda_{\alpha,3}}{m_{\alpha,3}}p_{\alpha,3}-\gamma M_z\sum_{\alpha=1}^{N}\frac{\lambda_{\alpha,1}}{m_{\alpha,1}}p_{\alpha,1} \nonumber \\
&+&\gamma M_zM_x\sum_{\alpha=1}^{N}\frac{\lambda_{\alpha,3}^2}{m_{\alpha,3}}-\gamma M_xM_z\sum_{\alpha=1}^{N}\frac{\lambda_{\alpha,1}^2}{m_{\alpha,1}},
\end{eqnarray}

\begin{eqnarray}
\frac{dM_x}{dt}&=&\gamma H_0M_y+\gamma M_z\sum_{\alpha=1}^{N}\frac{\lambda_{\alpha,2}}{m_{\alpha,2}}p_{\alpha,2}-\gamma M_y\sum_{\alpha=1}^{N}\frac{\lambda_{\alpha,3}}{m_{\alpha,3}}p_{\alpha,3} \nonumber \\
&+&\gamma M_yM_z\sum_{\alpha=1}^{N}\frac{\lambda_{\alpha,2}^2}{m_{\alpha,2}}-\gamma M_zM_y\sum_{\alpha=1}^{N}\frac{\lambda_{\alpha,3}^2}{m_{\alpha,3}}
\end{eqnarray}

\begin{equation}
\frac{dq_{\alpha,i}}{dt}=\frac{p_{\alpha,i}}{m_{\alpha},i}+\frac{\lambda_{\alpha,i}}{m_{\alpha,i}}M_i
\end{equation}

\begin{equation}
\frac{dp_{\alpha,i}}{dt}=-m_{\alpha,i}\omega_{\alpha,i}^2q_{\alpha,i}
\end{equation}

The linear equations (8) and (9) can easily be integrated and one can obtain the solution
\begin{eqnarray}
p_{\alpha,i}(t)&=&p_{\alpha,i}(0)\cos(\omega_{\alpha,i}t)-q_{\alpha,i}(0)m_{\alpha,i}\omega_{\alpha,i}\sin(\omega_{\alpha,i}t)\nonumber\\
&&-\lambda_{\alpha,i}\omega_{\alpha,i}\int_0^t dt^{\prime}\sin\lbrack\omega_{\alpha,i}(t-t^{\prime})\rbrack M_i(t^{\prime}),
\end{eqnarray}
where $p_{\alpha,i}(0)$ and $q_{\alpha,i}(0)$ are the initial momentums and positions of the bath variables respectively. In order to eliminate the bath variables we can substitute Eq. (10) in the equations of spin systems i.e. in Eqs. (5)-(7).  After substituting bath variables in spin system we obtain
\begin{eqnarray}
&&\frac{dM_z}{dt}=\gamma M_y\sum_{\alpha=1}^N\frac{\lambda_{\alpha,1}}{m_{\alpha,1}}\Big\lbrack -\lambda_{\alpha,1}\omega_{\alpha,1}\int_0^tdt^{\prime}\sin\lbrack\omega_{\alpha,1}(t-t^{\prime})\rbrack \ \nonumber \\ &&\times M_x(t^{\prime})+p_{\alpha,1}(0)\cos(\omega_{\alpha,1}t)-m_{\alpha,1}\omega_{\alpha,1}q_{\alpha,1}(0)\sin(\omega_{\alpha,1}t)
\Big\rbrack \nonumber \\
&&-\gamma M_x\sum_{\alpha=1}^N\frac{\lambda_{\alpha,1}}{m_{\alpha,1}}\Big\lbrack -\lambda_{\alpha,2}\omega_{\alpha,2}\int_0^tdt^{\prime}\sin\lbrack\omega_{\alpha,2}(t-t^{\prime})\rbrack \ \nonumber \\ &&\times M_y(t^{\prime})+p_{\alpha,2}(0)\cos(\omega_{\alpha,2}t)-m_{\alpha,2}\omega_{\alpha,2}q_{\alpha,2}(0)\sin(\omega_{\alpha,2}t)
\Big\rbrack \nonumber \\
&&+\gamma M_xM_y\sum_{\alpha=1}^N \frac{\lambda^2_{\alpha,1}}{m_{\alpha,1}}-\gamma M_yM_x\sum_{\alpha=1}^N \frac{\lambda^2_{\alpha,2}}{m_{\alpha,2}}
\end{eqnarray}
After doing integration by parts of the integral terms on the right hand side of Eq. (11) we obtain
\begin{eqnarray}
&&\frac{dM_z}{dt}=\gamma M_y\sum_{\alpha=1}^N\frac{\lambda_{\alpha,1}}{m_{\alpha,1}}\Big\lbrack \lambda_{\alpha,1}\int_0^tdt^{\prime}\cos\lbrack\omega_{\alpha,1}(t-t^{\prime})\rbrack \nonumber \\ &&\times\frac{dM_x(t^{\prime})}{dt^{\prime}}+p_{\alpha,1}(0)\cos(\omega_{\alpha,1}t)-m_{\alpha,1}\omega_{\alpha,1}q_{\alpha,1}(0)
 \nonumber \\
&&\times\sin(\omega_{\alpha,1}t)+\lambda_{\alpha,1}\cos(\omega_{\alpha,1}t)M_x(0)\Big\rbrack\nonumber \\
&&-\gamma M_x\sum_{\alpha=1}^N\frac{\lambda_{\alpha,2}}{m_{\alpha,2}}\Big\lbrack \lambda_{\alpha,2}\int_0^tdt^{\prime}\cos\lbrack\omega_{\alpha,2}(t-t^{\prime})\rbrack \frac{dM_x(t^{\prime})}{dt^{\prime}} \nonumber \\ &&+p_{\alpha,2}(0)\cos(\omega_{\alpha,2}t)-m_{\alpha,2}\omega_{\alpha,2}q_{\alpha,2}(0)\sin(\omega_{\alpha,2}t) \nonumber \\
&&+\lambda_{\alpha,2}\cos(\omega_{\alpha,2}t)M_y(0)\Big\rbrack
\end{eqnarray}
One can easily observe that Eq. (12) involves with transient terms involving the initial values of spin variables $M_x(0)$ and $M_y(0)$, One can easily make the argument that for long time behaviour,one can neglect these transient terms. We will come to this point in details later in this section. Similar kind of equations for $\frac{dM_x}{dt}$ and $\frac{dM_y}{dt}$ can be obtained with extra external magnetic field dependent terms $\gamma H_0M_y$  and $-\gamma H_0M_x$ respectively. One can easily write down the three equations for the three components of spin variable in a compact form as follows :
\begin{eqnarray}
\frac{dM_z}{dt}&=&\gamma M_y \eta_1(t)-\gamma M_x \eta_2(t)\nonumber \\
&&+\gamma M_y\int_0^t dt^{\prime}\Gamma_1(t-t^{\prime})\dot{M_x}(t^{\prime}) \nonumber \\
&&-\gamma M_x\int_0^t dt^{\prime}\Gamma_2(t-t^{\prime})\dot{M_y}(t^{\prime}),
\end{eqnarray}
Similarly for other components of the spin variable we can obtain
\begin{eqnarray}
\frac{dM_y}{dt}&=&-\gamma H_0M_x +\gamma M_x \eta_3(t)-\gamma M_z \eta_1(t)\nonumber \\
&&+\gamma M_x\int_0^t dt^{\prime}\Gamma_3(t-t^{\prime})\dot{M_z}(t^{\prime}) \nonumber \\
&&-\gamma M_z\int_0^t dt^{\prime}\Gamma_1(t-t^{\prime})\dot{M_x}(t^{\prime}),
\end{eqnarray}
and
\begin{eqnarray}
\frac{dM_x}{dt}&=&\gamma H_0M_y +\gamma M_z \eta_2(t)-\gamma M_y \eta_3(t)\nonumber \\
&&+\gamma M_z\int_0^t dt^{\prime}\Gamma_2(t-t^{\prime})\dot{M_y}(t^{\prime}) \nonumber \\
&&-\gamma M_y\int_0^t dt^{\prime}\Gamma_3(t-t^{\prime})\dot{M_z}(t^{\prime}),
\end{eqnarray}
where,
\begin{eqnarray}
\eta_i(t)&=&\sum_{\alpha=1}^{N}\frac{\lambda_{\alpha,i}}{m_{\alpha,i}}\Big\lbrack p_{\alpha,i}(0)\cos(\omega_{\alpha,i}t)-q_{\alpha,i}(0)m_{\alpha,i}\omega_{\alpha,i}\nonumber \\
&&\times\sin(\omega_{\alpha,i}t)\Big\rbrack
\end{eqnarray}
and
\begin{equation}
\Gamma_i(t-t^{\prime})= \sum_{\alpha=1}^{N} \frac{\lambda_{\alpha,i}^2}{m_{\alpha,i}}\cos\lbrack \omega_{\alpha,i}(t-t^{\prime})\rbrack
\end{equation}
Thus we obtain the required Langevin equations as mentioned in Eqs. (13)-(15). These equations involve with fluctuating terms $\eta_i(t)$ which contain initial values $(q_{\alpha,i}(0), p_{\alpha,i}(0))$ of the bath variables. We call these terms as fluctuating, since they involve with the uncertainty of the initial conditions of the bath variables. Therefore, at time $t=0$, it is customary to assume canonical equilibration of the heat bath oscillators at temperature $T$ with respect to the free oscillator Hamiltonian :
\begin{equation}
{\mathcal{H}}_B =\sum_{i=1}^{3} \sum_{\alpha=1}^{N} \Big\lbrack \frac{p_{\alpha,i}^2(0)}{2m_{\alpha,i}}+\frac{1}{2}m_{\alpha,i}\omega_{\alpha,i}^2q_{\alpha,i}^2(0)\Big\rbrack.
\end{equation}
Corresponding to this, the canonical distribution of the bath variables is given by
\begin{equation}
\rho_B=\frac{e^{-{\mathcal{H}}_B/k_BT}}{Z_B}; Z_B=\prod \int dq_{\alpha,i}(0)dp_{\alpha,i}(0)e^{-{\mathcal{H}}_B/k_BT}
\end{equation}
where, $k_B$ is the Boltzmann constant and $Z_B$ is the normalization factor, commonly known as partition function. The statistical average of a heat-bath variable,A, is given by
\begin{equation}
<A> = \frac{\int A e^{-{\mathcal{H}}_B/k_BT}}{Z_B}
\end{equation}
Now, using known properties of harmonic oscillator, it is straightforward to show that :
\begin{eqnarray}
<q_{\alpha,i}(0)>&=&0 \nonumber \\
<p_{\alpha,i}(0)>&=&0 \nonumber \\
<q_{\alpha,i}(0)q_{\beta,j}(0)>&=&\frac{k_BT}{m_{\alpha,i}\omega_{\alpha,i}^2}\delta_{i,j}\delta_{\alpha,\beta}, \nonumber \\
<p_{\alpha,i}(0)p_{\beta,j}(0)>&=&k_BT m_{\alpha,i}\delta_{i,j}\delta_{\alpha,\beta}, \nonumber \\
<p_{\alpha,i}(0)q_{\beta,j}(0)>&=&<q_{\alpha,i}(0)p_{\beta,j}(0)>=0,
\end{eqnarray}
In addition we have the Gaussian property : the statistical average of an odd number of factors of $q_{\alpha,i}(0)$ and $p_{\alpha,i}(0)$ is zero. On the other hand, even number of factors give us the sum of products of pair averages with the order of the factors preserved. Using the results in Eq. (21), one can show that the force $\eta_i(t)$ has zero mean,
\begin{equation}
<\eta_i(t)>=0
\end{equation}
and the correlation is given by
\begin{eqnarray}
<\eta_i(t)\eta_j(t^{\prime})>&=&k_BT\sum_{\alpha=1}^{N}\frac{\lambda_{\alpha,i}^2}{m_{\alpha,i}}\cos\Big\lbrack \omega_{\alpha,i}(t-t^{\prime})\Big\rbrack \nonumber \\
&=&\delta_{i,j}k_BT\Gamma_i(t-t^{\prime})
\end{eqnarray}
Thus, we obtain a relationship between the correlation of fluctuations and the memory kernels appear in Eqs. (13)-(15). This is actually known as fluctuation-dissipation theorem which is a must to preserve the thermodynamic consistency of a closed system. In addition, $\eta_i(t)$ has the Gaussian property, which follows from the same property of the $q_{\alpha,i}(0)$ and $p_{\alpha,i}(0)$. Thus the initial distribution of the heat bath oscillators makes the force $\eta_i(t)$ a Gaussian random force.\\
\subsection{ Markovian Limit}
The generalized Lanevin equations obtained in Eqs. (13)-(15) are in general non-Markovian and involve with the multiplicative fluctuating and concomitant nonlinear dissipative terms. But, one can easily obtain the Markovian limit of the Eqs. (13)-(15). This can be obtained by considering the memory kernel to be delta correlated, i.e., $\Gamma_i(t-t^{\prime})= 2\varsigma_i \delta(t-t^{\prime})$ which actually corresponds to the Ohmic dissipation \cite{caldeira}. For simplicity, we consider here strength of all $\varsigma_i$ to be equal which actually corresponds to the isotropic dissipation. Anisotrpic dissipation can also be readily treated in our scheme. But, the equations will be cumbersome and they can also be handled in studying the dynamics of the system (relaxation and equilibration). For the time being, we restrict our discussion only for the isotropic dissipation.  In this limit, one can easily show that Eqs. (13)-(15) actually reduces to that proposed by Kubo and Hashitsume $\frac{d\vec{M}}{dt}= \gamma \lbrack \vec{H}_0+\vec{H}(t)-\varsigma\frac{d\vec{M}}{dt}\rbrack \times \vec{M} $ with $\vec{H}_0=\hat{k}H_0$ and $\vec{H}(t)= \hat{i}\eta_1(t)+\hat{j}\eta_2(t)+\hat{k}\eta_3(t)$. The unit vectors in the x,y and z directions are denoted by $\hat{i}$, $\hat{j}$ and $\hat{k}$, respectively. It is to be noticed that the neglected transient terms in Eq. (12) can be represented as $M_x(t)M_y(0)\delta(t)$ and $M_y(t)M_x (0)\delta(t)$ which confirms that the transient terms can safely be neglected as they don't contribute in the long time behaviour of the spin relaxation. It is a cautious remark to the reader that the transient terms should not be treated ideally as delta functions in the Markovian limit. Otherwise, a jump problem in the initial condition may arise. For example, in the Markovian limit or Ohmic case \cite{caldeira}, the spectral density for the bath oscillators is usually chosen as $J(\omega)= \frac{\pi}{2}\sum_{\alpha}\frac{\lambda_{\alpha}^2}{m_{\alpha}\omega_{\alpha}}\delta(\omega-\omega_{\alpha})=\varsigma\omega$ with a sharp cut-off. This kind of spectral density has no upper bound which is nonphysical. Thus, one should introduce an upper cut-off frequency $\omega_c$, e.g., $J(\omega)=\varsigma\omega e^{-\omega/\omega_c}$. This frequency scale $\omega_c$ is usually much larger than all other characteristic frequencies of the system. Thus, the Markovian limit corresponds to the time scale $t>>1/\omega_c$ and this can be made arbitrarily small. So, transient terms survive up to the time scale $1/\omega_c$ and one should consider transient terms as functions with a finite but arbitrarily small width in the Markovian limit. For all other non-Markovian forms of the spectral density of the heat bath one must include the neglected transient terms in the problem for the full dynamical evolution.
\subsection{Seshadri-Lindenberg Equations}
One can easily derive the dissipative equations obtained by Seshadri and Lindenberg \cite{seshadri} from our Eqs. (13)-(15). As we have shown earlier that we can obtain Kubo-Hashitsume dissipative equation  $\frac{d\vec{M}}{dt}= \gamma \vec{M}\times\lbrack \vec{H}_0+\vec{H}(t)-\varsigma\frac{d\vec{M}}{dt}\rbrack $ with $\vec{H}_0=\hat{k}H_0$ and $\vec{H}(t)= \hat{i}\eta_1(t)+\hat{j}\eta_2(t)+\hat{k}\eta_3(t)$ from Eqs. (13)-(15) in the Markovian limit. As a perturbation, if one replace the term $\frac{d\vec{M}}{dt}$ by the evolution term in the absence of coupling, i.e., by $\gamma\lbrack \vec{H}_0\times \vec{M}\rbrack$, one can arrive at :
\begin{equation}
\frac{d\vec{M}}{dt}= \gamma\vec{M}\times\lbrack \vec{H}_0+\vec{H}(t)-\varsigma\gamma(\vec{M}\times\vec{H}_0)\rbrack
\end{equation}
This is a well known form of Landau-Lifshitz dissipative equation. In addition, if we assume isotropic fluctuation,i.e., $\eta_1(t)= \eta_2(t)=\eta_3(t)= \eta(t)$ , we can obtain from Eq. (24) :
\begin{eqnarray}
\frac{dM_z}{dt}&=& \frac{i}{2}\lbrack h(t)M_-(t)-h^*(t)M_+(t)\rbrack +H_0\varsigma M_+M_-, \nonumber \\
\frac{dM_+}{dt}&=& i\Omega M_+(t)-ih(t)M_z(t)-H_0\varsigma M_zM_+, \nonumber \\
\frac{dM_-}{dt}&=& -i\Omega M_-(t)+ih^*(t)M_z(t)-H_0\varsigma M_zM_-,
\end{eqnarray}
where, we have used
\begin{eqnarray}
\Omega=-H_0+\eta(t) , \nonumber \\
h(t)= \eta(t)+i\eta(t); h^*(t)= \eta(t)-i\eta(t), \nonumber \\
M_+=M_x+iM_y ; M_-=M_x-iM_y,
\end{eqnarray}
and we take $\gamma=1$. Equations obtained in (25) are same as that of dissipative equations obtained by Seshadri and Lindenberg. Only difference is observed in $\Omega$. Unlike, Seshadri and Lindenberg our $\Omega$ explicitly depends on the fluctuatin force $\eta(t)$.
\section{Fokker Planck Equation}
 In this section, we derive the Fokker Planck Equation corresponding to our generalized Langevin Eqs. (14)-(16). Here, we adopt the method introduced in Refs. \cite{garcia1,garcia2}. It is to be remembered that our nonlinear generalized Langevin Eqs. (13)-(15) can be cast in the Markovian limit into the following form:
 \begin{equation}
  \frac{d\vec{M}}{dt}=\gamma\vec{M}\times\lbrack \vec{H_0}+\vec{H}(t)\rbrack -\gamma\zeta\vec{M}\times\frac{d\vec{M}}{dt} ,
 \end{equation}
  with $\vec{H_0}=\hat{k}H_0$ and $\vec{H}(t)=\hat{i}\eta_1(t)+\hat{j}\eta_2(t)+\hat{k}\eta_3(t)$. Equation (27) involves a term proportional to $\frac{d\vec{M}}{dt}$ on its right hand side, henceforth it can be called as stochastic Gilbert type equation. In the weak damping limit ($\eta<<1$), one can show that Eq. (27) reduces to stochastic Landau-Lifshitz form, i.e., Eq. (24). In general the stochastic general Langevin equations are defined as :
  \begin{equation}
    \frac{dy_i}{dt}= A_i(\vec{y},t)+ \sum_k B_{i,k}(\vec{y},t)\xi_k(t),
  \end{equation}
 where, $\vec{y}=(y_1,y_2,....y_n)$ is a multicomponent stochastic process, $k$ runs over a given set of indices and Langevin source $\xi_k(t)$ are independent Gaussian stochastic process which satisfies :
 \begin{eqnarray}
 <\xi_k(t)>=0 \nonumber \\
  <\xi_k(t)\xi_l(s)>=2D\delta_{k,l}\delta(t-s).
  \end{eqnarray}
  When the functions $B_{i,k}(\vec{y},t)$ depend on $\vec{y}$ the noise is termed as multiplicative noise, while $\frac{\partial B_{i,k}}{\partial y_j}=0$ defines the additive noise. Thus, the nonequilibrium Probability distribution function of $\vec{y}$ at time $t$,i.e., $P(\vec{y},t)$ obeys the following Fokker Planck Equation \cite{garcia1,garcia2}:
 \begin{eqnarray}
 &&\frac{\partial P}{\partial t}=-\sum_i \frac{\partial}{\partial y_i}\Big \lbrack \lbrace A_i(\vec{y},t)+D\sum_{j,k} B_{j,k}(\vec{y},t)\frac{\partial B_{i,k}(\vec{y},t)}{\partial y_j}\rbrace  \nonumber \\
 &&\times P\Big\rbrack+D\sum_{i,j}\frac{\partial^2}{\partial y_i \partial y_j}\Big\lbrack \lbrace\sum_k B_{i,k}(\vec{y},t)B_{j,k}(\vec{y},t)\rbrace P \Big \rbrack
 \end{eqnarray}
 The stochastic Landau-Lifshitz-Gilbert equation can be cast into the general form of Langevin equation (Eq. 28) with the help of the statistical properties in Eq. (29) by identifying $(y_1,y_2,y_3)= (M_x,M_y,M_z)$, $\xi_k(t)=H(t)$,
 \begin{eqnarray}
 A_i&=&\sum_{j,k}\epsilon_{i,j,k}M_jH_{0k}+\lambda \sum_k (M^2\delta_{i,k}-M_iM_k)H_k \nonumber \\
 B_{i,k}&=&\sum_j \epsilon_{i,j,k}M_j,
 \end{eqnarray}
 where, $\epsilon_{i,j,k}$ are antisymmetrical unit tensor of rank three (Levi-Civita symbol), $\lambda=\gamma\zeta$ and we have expanded the vector triple product $-\vec{M}\times(\vec{M}\times\vec{H}_0)$ by using the rule $a\times (b\times c)=\vec{b}(\vec{a}.\vec{c})-\vec{c}(\vec{a}.\vec{b})$. To calculate the noise induced drift term of the Fokker Planck equation we require the derivative of the diffusion coefficient \cite{garcia1,garcia2} :
 \begin{equation}
   \frac{\partial B_{i,k}}{\partial M_j}=\epsilon_{i,j,k}.
 \end{equation}
 Thus,
 \begin{eqnarray}
 D \sum_{j,k}B_{j,k}\frac{\partial B_{i,k}}{\partial M_j} &=& D\sum_l(\sum_{j,k}\epsilon_{jlk}\epsilon_{ijk})M_l \nonumber\\
    &=& -2D\sum_{l}\delta_{il}M_l=-2DM_i .
 \end{eqnarray}
 Let us compute the coefficient $\sum_{k}B_{ik}B_{jk}$ \cite{garcia1,garcia2}
 \begin{eqnarray}
   \sum_{k}B_{ik}B_{jk} &=& \sum_k \lbrack\sum_r \epsilon_{irk}M_r\rbrack \lbrack \sum_s \epsilon_{jsk}M_s\rbrack  \nonumber \\
    &=& \sum_{r,s}M_rM_s \Big\lbrack \sum_k \epsilon_{irk}\epsilon_{jsk}\Big\rbrack \nonumber \\
    &=& \sum_{r,s}M_rM_s\lbrack \delta_{i,j}\delta_{r,s}-\delta_{i,s}\delta_{r,j}\rbrack \nonumber \\
    &=& M^2\delta_{i,j}- M_iM_j
 \end{eqnarray}
 Thus, the corresponding Fokker-Planck equation becomes \cite{garcia1,garcia2}
 \begin{eqnarray}
   \frac{\partial P}{\partial t} &=& -\frac{\partial}{\partial M_i}\Big\lbrack \Big\lbrace \sum_{j,k}\epsilon_{ijk}M_jH_{0k}+\lambda \sum_k(M^2\delta_{i,k}-M_iM_k)H_{0k}  \nonumber \\
    &-&2DM_i\Big\rbrace P\Big\rbrack + D \sum_{ij}\frac{\partial^2}{\partial M_i\partial M_j}\Big\lbrack (M^2\delta_{ij}-M_iM_j)P\Big\rbrack \nonumber \\
 \end{eqnarray}
 Now, taking the $M_j$ derivative of the last term using the fact $\sum_j\partial_i\Big(M^2\delta_{i,j}-M_iM_j \Big)=\sum_j (2M_j \delta_{i,j}-M_i\delta_{j,j}-M_j\delta_{i,j}= 2M_i-3M_i-M_i=-2M_i$. Thus, we obtain from the last term $D\sum_i\partial_i\Big\lbrack -2M_iP+\sum_{j}(M^2\delta_{i,j}-M_iM_j)\partial_jP$. Here, the first term cancel with the $-2DM_iP$ term of the drift and the last term can be combined with $\sum_k(M^2\delta_{i,k}-M_iM_k)H_{0k}$. Finally, the Fokker Planck equation in vector notation becomes :
 \begin{equation}
 \frac{\partial P}{\partial t}= -\frac{\partial }{\partial\vec{M}}. \Big\lbrack \vec{M}\times \vec{H}_0-\vec{M}\times\Big\lbrace \vec{M}\times \Big(\lambda\vec{H}_0-D\frac{\partial}{\partial\vec{M}}\Big)\Big\rbrace\Big\rbrack P
 \end{equation}
 \subsection{Stationary Solution}
 In order to ensure that our system's stationary properties coincide with the appropriate thermal equilibrium properties, the Fokker-Planck equation must have the Boltzmann distribution :
 \begin{equation}
   P_{0}(\vec{M})= \frac{1}{Z}\exp\Big(-\beta\mathbf{H}(\vec{M})\Big)
 \end{equation}
 as a stationary solution. Since, we can write $\vec{H}_0=- \frac{\partial\mathbf{H}(\vec{M})}{\partial \vec{M}}$, we can write
 \begin{equation}
 \frac{\partial P_0}{\partial \vec{M}}=\beta \vec{H}_0P_0.
 \end{equation}
 This implies $\vec{M}\times \vec{H}_0P_0$ is divergenceless. Thus, using these results, in order to have the Boltzmann distribution as stationary solution of the Fokker-Planck equation, one should set $D\beta = \lambda$, i.e., $D=\lambda k_BT$.\\
 \subsection{Longitudinal and Transverse components}
In this subsection we discuss about the longitudinal and transverse components of the  magnetic spin of our test particle. It is possible to separate out the longitudinal and transverse dynamics of the magnetic moment associated with the test particle under certain limiting conditions. As we have already discussed that our Langevin equation can be cast into a Kubo and Hashitsume form in the Markovian limit (Eq. 27). Now, if we insert again $d\vec{M}/dt$ from Eq. (27) in the third term of Eq. (27) we obtain :
\begin{equation}
(1+\zeta^2\gamma)\frac{d\vec{M}}{dt}=\gamma\vec{M}\times\lbrack \vec{H_0}+\vec{H}(t)\rbrack -\gamma\zeta\vec{M}\times\lbrack\vec{M}\times\vec{H_0}+\vec{H}(t)\rbrack
\end{equation}
Projecting Eq. (39) along the unit Magnetization vector $\hat{M}=\frac{\vec{M}}{M}$ we obtain
\begin{equation}
\frac{dM_{\parallel}}{dt}=0
\end{equation}
Consequently one can show that $\vec{M}\cdot\frac{d\vec{M}}{dt}=0$ which implies the magnitude of the magnetization vector is conserved. The corresponding Fokker Planck equation becomes $\partial_tP(M,t)= -\partial_Mj$ with $j= -D\partial_M P+\eta H_{\parallel}$ and $H_{\parallel}=\vec{H}\dot\hat{M}$. Now, setting the dissipative current $j=0$ one can obtain the detailed balance equilibrium distribution $P_{eq}(M)= P_0\exp(-g_{\parallel}V/k_BT)$ with energy density $g_{\parallel}= -H_{\parallel}M$. Now, subtracting longitudinal Eq. (40) from Eq. (39) we obtain the transverse dynamics
\begin{equation}
\frac{dM_{\perp}}{dt}=\gamma^{\prime}\vec{M}\times\lbrack (\vec{H}_0+\vec{H}(t))-\zeta^{\prime}\vec{M}\times(\vec{H}_0+\vec{H}(t))\rbrack,
\end{equation}
where $\vec{M}_{\perp}=\vec{M}-\hat{M}M$, $\gamma^{\prime}=\frac{\gamma}{1+\zeta^2\gamma^2}$, and $\zeta^{\prime}=\frac{\zeta\gamma}{1+\zeta^2\gamma^2}$. One can note that for this constant magnitude $M$, $d\vec{M}_{\perp}/dt=d\vec{M}/dt$  and Eq. (41) is basically the same stochastic LLG equation as that of Eq. (39). Thus, the equilibrium distribution will also be same as that of Eq. (37).
 \section{Non-Markovian System:}
In the Non-Markovian regime one can not simply drop the terms involving initial values $\Big(M_x(0),M_y(0),M_z(0)\Big)$ which usually is simply brushed under the rug for strict Ohmic case. For strict Ohmic dissipation they reduce to $\delta$ function contributions and onecan easily drop them. But, in the Non-Markovian limit we can not drop these terms.  Thus, Eqs. (13)-(15) are modified as follows :
\begin{eqnarray}
\frac{dM_z}{dt}&=&\gamma M_y \lbrack\eta_1(t)+\Gamma_1(t)M_x(0)\rbrack -\gamma M_x\lbrack\eta_2(t)\nonumber \\
&&+\Gamma_2(t)M_y(0)\rbrack+\gamma M_y\int_0^t dt^{\prime}\Gamma_1(t-t^{\prime})\dot{M_x}(t^{\prime}) \nonumber \\
&&-\gamma M_x\int_0^t dt^{\prime}\Gamma_2(t-t^{\prime})\dot{M_y}(t^{\prime}),
\end{eqnarray}

\begin{eqnarray}
\frac{dM_y}{dt}&=&-\gamma H_0M_x +\gamma M_x\lbrack \eta_3(t)+\Gamma_3(t)M_z(0)\rbrack\nonumber \\
&-&\gamma M_z \lbrack\eta_1(t)+\Gamma_1(t)M_x(0)\rbrack
+\gamma M_x\int_0^t dt^{\prime}\Gamma_3(t-t^{\prime}) \nonumber \\
&&\times\dot{M_z}(t^{\prime})-\gamma M_z\int_0^t dt^{\prime}\Gamma_1(t-t^{\prime})\dot{M_x}(t^{\prime}),
\end{eqnarray}
and
\begin{eqnarray}
\frac{dM_x}{dt}&=&\gamma H_0M_y +\gamma M_z \lbrack\eta_2(t)+\Gamma_2(t)M_y(0)\rbrack\nonumber \\
&-&\gamma M_y \lbrack\eta_3(t)+\Gamma_3(t)M_z(0)\rbrack
+\gamma M_z\int_0^t dt^{\prime}\Gamma_2(t-t^{\prime}) \nonumber \\
&&\times\dot{M_y}(t^{\prime})-\gamma M_y\int_0^t dt^{\prime}\Gamma_3(t-t^{\prime})\dot{M_z}(t^{\prime}),
\end{eqnarray}
It is now customary to introduce an auxiliary random force $\zeta_k(t)=\eta_k(t)+\Gamma_k(t)M_k(0); k=1,2,3$. In terms of this new random force the  Langevin Like Equations (39)-(41) no longer assume the form of GLE. Although they look like GLE. They now contain initial inhomogeneous slip term $\Gamma_k(t)M_k(0)$. The combined noise force,$\zeta_i(t)=$, :
\begin{eqnarray}
\zeta_i(t)&=&\sum_{\alpha=1}^{N}\frac{\lambda_{\alpha,i}}{m_{\alpha,i}}\Big\lbrack (p_{\alpha,i}(0)+\lambda_{\alpha,i}M_i(0))\cos(\omega_{\alpha,i}t)\nonumber \\
&&-q_{\alpha,i}(0)m_{\alpha,i}\omega_{\alpha,i}\sin(\omega_{\alpha,i}t)\Big\rbrack
\end{eqnarray}
is no longer has a stationary autocorrelation when averaged with respect to the bare bath ensemble as introduced in Eq (19) in combination with Eq. (18). The combined stochastic force $\zeta_i(t)$ can be stationary and colored Gaussian, if we consider a conditional average with respect to the following Gaussian equilibrium ensemble :
\begin{eqnarray}
&&\rho_i\Big(\lbrace p_{\alpha,i},q_{\alpha,i}\rbrace | M_i(t_0)=M_i(0)\Big) =\frac{1}{Z}\nonumber \\
&&\exp\Big\lbrack -\beta \Big\lbrace \sum_{\alpha}\frac{(p_{\alpha,i}+\lambda_{\alpha,i}M_i(0))^2}{2m_{\alpha,i}}
   +\frac{m_{\alpha,i}\omega_{\alpha,i}^2q_{\alpha,i}^2}{2}\Big\rbrace\Big\rbrack \nonumber \\
\end{eqnarray}
Considering this conditional probability for the bath variables we can again obtain identical fluctuation dissipation relations for the stochastic force $\zeta_i(t)$ as that of $\eta_i(t)$ as mentioned in Eqs. (22) and (23):
\begin{eqnarray}
  <\zeta_i(t)>_{\rho_i} &=& 0 \\
   <\zeta_i(t)\zeta_j(s)>_{\rho}&=& \delta_{i,j}k_BT\Gamma_i(t-s)
\end{eqnarray}
This will again enables us to obtain the same kind of equilibrium distribution as that of Eq. (37), but with respect to a different Gaussian ensemble and colored noise force.

\section{Conclusions}
We discuss the relaxation of a single classical spin interacting with a heat bath through momentum variables in the framework of GLE through dynamical model Hamiltonian description. The basic features of this momentum dissipative spin model are found to be identical as that of discussed in Ref. \cite{jayannavar} for a spin model with coordinate-coordinate coupling. The Fokker-Planck equations  corresponding to the GLE for our momentum dissipative spin model are also derived in the Markovian as well as non-Markovian limit. The stationary solutions of these Fokker-Planck equations are also analyzed. In the non-Markovian limit, the difficulty to obtain the stationary solution of the Fokker-Planck equation is discussed in details. To overcome this, one can introduce an auxiliary random force and can obtain the stationary solution.
\begin{acknowledgments}
MB acknowledge the financial support of IIT Bhubaneswar through seed money project SP0045. AMJ thanks DST, India for financial support.
\end{acknowledgments}

\end{document}